# Magnetization "Steps" on a Kagome Lattice in Volborthite

Hiroyuki Yoshida, Yoshihiko Okamoto, Takashi Tayama, Toshiro Sakakibara, Masashi Tokunaga, Akira Matsuo, Yasuo Narumi, Koichi Kindo, Makoto Yoshida, Masashi Takigawa and Zenji Hiroi*

*Institute for Solid State Physics, University of Tokyo, Kashiwa, Chiba 277-8581, Japan*

Magnetic properties of the spin-1/2 kagome-like compound volborthite are studied using a high-quality polycrystalline sample. It is evidenced from magnetization and specific heat measurements that the spins on the kagome lattice still fluctuate at low temperature, down to $T = 60$ mK that corresponds to 1/1500 of the nearest-neighbor antiferromagnetic interaction, exhibiting neither a conventional long-range order nor a spin gap. In contrast, $^{51}$V NMR experiments revealed a sharp peak at 1 K in relaxation rate, which indicates that a certain exotic order occurs. Surprisingly, we have observed three "steps" in magnetization as a function of magnetic field, suggesting that at least four liquid-like or other quantum states exist under magnetic fields.

KEYWORDS: kagome lattice, spin liquid, magnetization step, volborthite

*E-mail address: hiroi@issp.u-tokyo.ac.jp

The magnetic properties of materials arise from the collective interaction of electron spins on atoms in a crystal. The antiferromagnetic interaction often causes long-range order (LRO) in an alternating up-down pattern called the Néel order at low temperatures. However, such LRO can be destroyed completely on trianglular lattices, because the antiferromagnetic interaction is inevitably frustrated on each triangle, suppressing a unique arrangement of spins covering the whole lattice.[1,2] It is expected therefore that the system remains "liquid", called the spin liquid, instead of a Néel order down to the lowest temperature. Since quantum fluctuations should play a crucial role in such a quantum disordered state, one would expect a new state of matter with properties we have never encountered.[3] Moreover, a quantum spin liquid might have exotic excitations or adopt a certain type of subtle order, such as a topological order.[4] Although an extensive study has been carried out to search for this mysterious state, clear experimental evidence has remained elusive until recently.[3,4]

One famous candidate for a quantum spin liquid is a resonating-valence-bond (RVB) state on a triangular lattice postulated theoretically by Anderson in 1973.[5] Instead of an ordinary Néel state with LRO, the RVB state consists of paired spins with zero total quantum spin number, such as the configurations on a kagome lattice depicted in Fig.1(a). Since the ground state is defined quantum mechanically as a linear combination of all possible configurations derived from different pairings, a liquid-like behavior is expected even at $T = 0$.[6]

A kagome lattice made of vertex-sharing triangles is one of the typical playgrounds for frustration physics.[7] The theoretical ground state for the spin-1/2 kagome antiferromagnet (KAFM) is in fact an RVB state with an energy gap called the spin gap in the excitation spectrum.[1,8,9] This is because in the RVB picture, an excitation results from breaking spin singlet pairs: the larger the magnetic coupling, the larger the spin gap. Since the predicted magnitude of the spin gap $\Delta$ is small, $J/4$ or $J/20$, where $J$ is the magnitude of the nearest-neighbour antiferromagnetic coupling, one assumes that there exists a long-range RVB state including a wide range of extended singlet pairs with smaller coupling constants for larger separations, instead of a short-range RVB state having $\Delta \sim J$ made of all local singlet pairs. It seems difficult, however, to determine the magnitude of the gap, because of the inherent difficulty in carrying out theoretical calculations in the presence of frustration.

A challenge for materials scientists is to search for a suitable natural material to realize this exotic state of matter.[10] Despite intensive efforts, experimental confirmations of the spin liquid have been scarce and sometimes controversial:[10-17] most candidate materials suffer from more or less disorder arising from crystalline defects or impurities and unwanted anisotropic or three-dimensional interactions.[3,4] Volborthite $Cu_3V_2O_7(OH)_2 \cdot 2H_2O$ is a two-dimensional antiferromagnet comprising $Cu^{2+}$ ions arranged in a kagome-like lattice[18] and has been presumed to form a spin liquid on the kagome lattice without LRO down to 50 mK.[19-22] However, the recent study conducted by Bert *et al.* revealed glass-like freezing at 1.2 K as well as a mixture of different spin configurations,[23] suggesting that the ground state of volborthite is also far from the true spin liquid state. The polycrystalline samples used in the previous study were prepared by precipitation and typically contained 1% of almost noninteracting spins in the kagome plane.[19,21-24] Recently, we have tried to improve the sample quality and succeeded in preparing a sample of much higher quality; this was found to contain less than 0.1% of free spins, more than one order smaller than the previous value. This large reduction in the number of free impurity spins has enabled us to investigate here the intrinsic properties of the kagome lattice in volborthite.

A polycrystalline sample was prepared by precipitation described previously.[18,19] To improve its quality, the sample was annealed under hydrothermal conditions at 190°C for 12 h after precipitation. Powder X-ray diffraction profiles become much sharper after annealing, displaying marked improvements in crystallinity and particle size. Magnetization at moderately high magnetic fields was measured in a Quantum Design MPMS equipment between 2 and 350 K and in a Faraday-force capacitive magnetometer down to 60 mK.[25] High-field magnetization measurements were carried out using a pulsed magnet up to 55 T at $T$ = 1.4 and 4.2 K. Specific heat was measured in a Quantum Design PPMS equipment down to 0.5 K.

The magnetic susceptibility $\chi$ shown in Fig. 1(a) in a wide temperature range exhibits a Curie-Weiss increase on cooling from high temperature, followed by a broad maximum at $T_p \sim$ 22 K without any anomaly indicating LRO. From fitting to the theoretical model for the $S$-1/2 KAFM[8] above 150 K, the average antiferromagnetic interaction[26] is determined to be $J$ = 86 K on the basis of the spin Hamiltonian $J\sum S_i \cdot S_j$. Thus, $T_p$ corresponds approximately to $J/4$.[24] Another marked indication from the $\chi$ data is the absence of any spin-gap behavior. Although $\chi$ would tend to zero as $T$ approaches zero, if a gap is opened, as illustrated in Fig. 1(a) for the theoretical case of $\Delta = J/4$,[8] the $\chi$ of volborthite can remain large and finite at $\sim 3 \times 10^{-3}$ cm$^3$ mol-Cu$^{-1}$, implying the absence of a gap or the presence of a very small gap. Furthermore, we have extended our $\chi$ measurements down to 60 mK, as shown in Fig. 1(b), and observed an almost $T$-independent behavior with neither an anomaly nor any indication of a downturn. Therefore, the spin gap can be no more than $J/1500$, which is much smaller than theoretically predicted values.[8,9] This strongly suggests that the ground state of volborthite is nearly gapless and probably a spin liquid.

Spin glass transitions are observed even in our clean sample at $T_g$ = 1.1 and 0.32 K at magnetic fields of 0.1 and 1 T, respectively (Fig. 1(b)). It has been pointed out, however, based on the previous NMR results, that this spin glass can be associated with domains based around impurity spins having local staggered moments and, therefore, is not intrinsic.[19,24] Fortunately, because the impurity-induced spin glass disappears with increasing field, we can study the intrinsic properties of the kagome lattice at high magnetic fields, above 2 T.

Microscopic probes, such as μSR and NMR, are frequently used to investigate the dynamics of spins. The previous μSR study[21,22] revealed a significant increase in relaxation rate $\lambda$ at low temperatures below 3 K, towards $T \sim$ 1 K (Fig. 2(a)), due to the slowing down of the spin fluctuations, which remain dynamic with a correlation time of 20 ns down to 50 mK.[27] On the other hand, we observed a sudden broadening of the $^{51}$V NMR line below 1 K, as shown in the inset of Fig. 2(a). Such a broadening is usually considered as evidence of static magnetic order. More precisely, a certain fraction of spin moments must be static on the time scale of the spin-echo measurements (10 - 50 μs). Correspondingly, the NMR relaxation rate $1/T_1$ shows a sharp peak at $T^* \sim 1$ K due to the critical slowing down of spin fluctuations (Fig. 2(a)). Note that the measurements were carried out at a magnetic field of 1 T, where a weak spin glass transition can be detected at the lower temperature of 0.32 K (Fig. 2(b)). It is also pointed out that the peak at $1/T_1$ remains at nearly the same temperature at a higher magnetic field of 4 T. In contrast, the truly static thermodynamic quantities, such as $\chi$ and heat capacity $C_p$, do not exhibit any anomaly indicative of a phase transition at $T^*$, as shown in Fig. 2(b). This contrasting behavior among μSR, NMR and static measurements indicates a specific dual character of the collective ground state. It is plausible that unusually slow fluctuations persist down to $T = 0$, without a completely static spin order, realizing a sort of spin liquid ground state. Alternatively, although a certain degree of order (hidden order) occurs below $T^*$, the order parameter does not couple to the magnetization measurements.

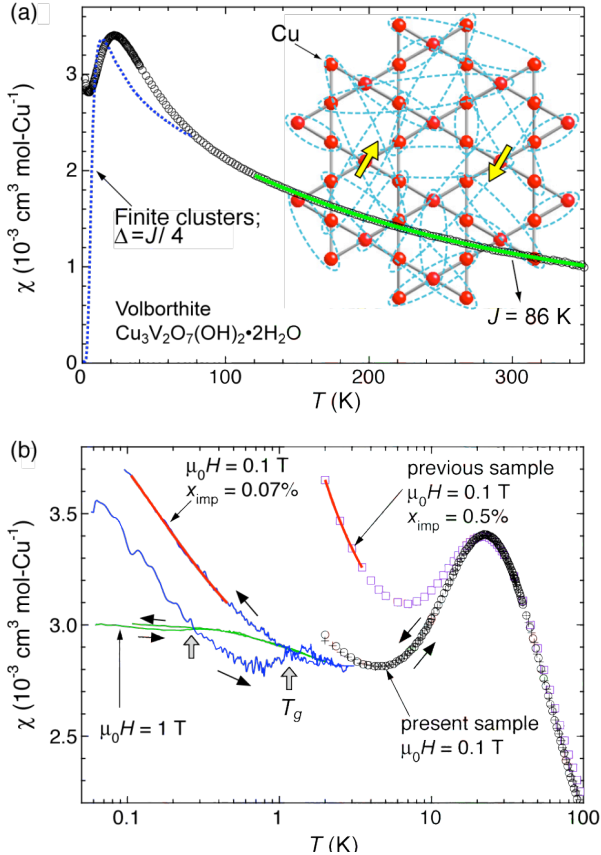

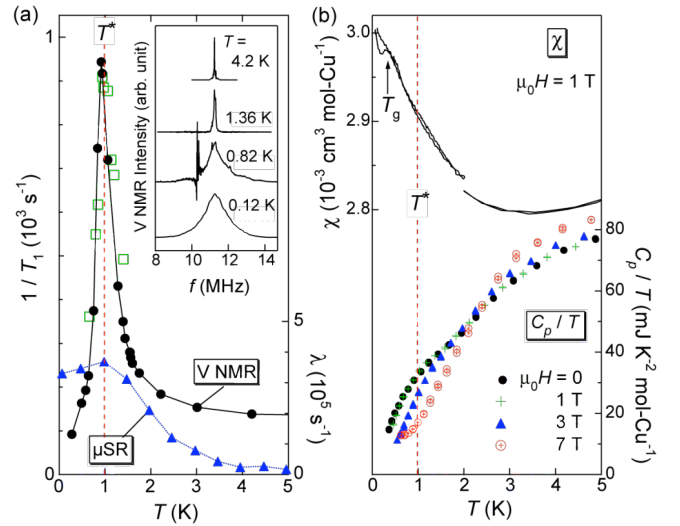

Fig. 1. Temperature dependence of the magnetic susceptibility $\chi$ of volborthite measured using a high-quality polycrystalline sample. (a) $\chi$ for the wide temperature range measured with a Quantum Design MPMS at $\mu_0H$ = 0.1 T on heating, after cooling at zero field. The solid curve above 150 K represents a fit to the theoretical model for the $S$-1/2 kagome antiferromagnet,[8] which yields $J$ = 86 K. The dotted curve is obtained from theoretical calculations on finite clusters for a spin gap of $\Delta = J/4$ to open.[8] The inset schematically shows a snapshot of a long-range resonating-valence-bond state on a kagome lattice made of Cu atoms shown by balls, which consists of various ranges of spin-singlet pairs, as indicated by broken ovals. (b) $\chi$ measured by the Faraday method with a dilution refrigerator on heating from 60 mK and cooling to 100 mK at $\mu_0H$ = 0.1 and 1 T. The $\chi$ data obtained above 2 K from a previous, low-quality sample are also shown for comparison. The thick curve on each dataset is a fit to the Curie-Weiss law, indicating that the concentrations of nearly free spins $x_{imp}$ are 0.5 and 0.07% for the previous and present samples, respectively.

Fig. 2. (a) Relaxation rates $\lambda$ (triangles) from previous μSR measurements[21] at $\mu_0H$ = 0.01 T and $1/T_1$ from the present $^{51}$V NMR experiments at $\mu_0H$ = 1 (circles) and 4 (squares) T. The inset shows the temperature evolution of the NMR spectra taken at $\mu_0H$ = 1 T at frequencies between 8 and 14.5 MHz. (b) Magnetic susceptibility $c$ measured at $\mu_0H$ = 1 T (the same data given in Fig. 1(a)) and heat capacity divided by temperature $C_p/T$ at $\mu_0H$ = 0, 1, 3 and 7 T obtained in a Quantum Design PPMS system.

Surprisingly, we have observed an unusual behavior in the M-H curve at low temperature and after the suppression of the spin glass, as shown in Fig. 3, which was not detected in previous samples, but has shown up as a result of a distinct improvement in sample quality. At first glance, the M-H curves appear to be proportional to H. However, there is a well-defined peak in the derivative curve at $T = 60$ mK and $\mu_0 H_{s1} = 4.3$ T, which indicates a step-like increase in M at $H_{s1}$. To make it discernible, an initial H-linear component has been subtracted, and the remaining $\Delta M$ is plotted at the bottom of Fig. 3(a). Obviously, the M-H curve starts to deviate from the initial linear relation at ~2 T and increases rather steeply above 4 T in an S-shaped curve. On heating, the increase appears to fade away at the same field. This step-like increase in M reminds us of a spin-flop transition in an ordered antiferromagnet or a metamagnetic transition in a magnetic alloy. However, because there is no conventional antiferromagnetic order in volborthite, it must be something else.

Even more surprisingly, we found more "magnetization steps" at higher fields, as shown in Fig. 3(b). Similar step-like or S-shaped increases are observed at $\mu_0 H_{s2} = 25.5$ T and $\mu_0 H_{s3} = 46$ T. It is striking that the M values at these fields, where the curves measured at $T = 1.4$ and 4.2 K cross each other, correspond to 1/6 and 1/3 of the saturation magnetization $M_s = 1$ $\mu_B$ for $S = 1/2$, where $\mu_B$ is the Bohr magneton. In this context, the first anomaly at $\mu_0 H_{s1} = 4.3$ T has $M \sim M_s / 45$. It is interesting, moreover, that the interval between $H_{s1}$ and $H_{s2}$ is nearly equal to that between $H_{s2}$ and $H_{s3}$, approximately 21 T, corresponding to $J / 6$. Again, in this context, $H_{s1}$ corresponds to $J / 30$. Since we observed a similar M-H curve at a lower temperature of 100 mK, the broad transitions observed cannot be due to thermal effects. In fact, the transition is still broad even at 60 mK in the first step. One possibility is that this is caused by a small distribution in the transition fields depending on the field directions, because a powder sample containing randomly oriented particles was used in the measurements.

One might observe real steps in M on a single crystal. Nevertheless, it is likely that the real steps are too small to observe, because they would be associated with transitions or crossover from one quantum spin liquid to another, with no symmetry breaking.

The origin of these magnetization steps is not clear. They are substantially different from other well-known magnetization plateaus, where a certain state with a fractional magnetization is stabilized over a field range.[28, 29] In a strong contrast, such fractional states are destabilized in the magnetization steps observed here. The coincidences of the field intervals with $J / 30$ or $J / 6$ suggest that there is a simple energy scheme in the excitation spectrum. One primitive speculation as a starting point is that the system changes from a long-range RVB to a short-range RVB state with increasing field, because the magnetic field may effectively reduce the correlation length or, in other words, diminish the number of widely separated singlet pairs by first creating triplet pairs: the number of excited triplets in a singlet "sea" corresponds to the magnitude of the magnetization in the RVB picture. Nevertheless, one may also take into account the effects of deviations from the simple KAFM model present in the real material of volborthite, such as the slight deformation of the kagome lattice or complex interactions other than the Heisenberg type. It seems rather difficult to explain these simple-looking phenomena within our present knowledge of the physics of the KAFM or other quantum spin models.

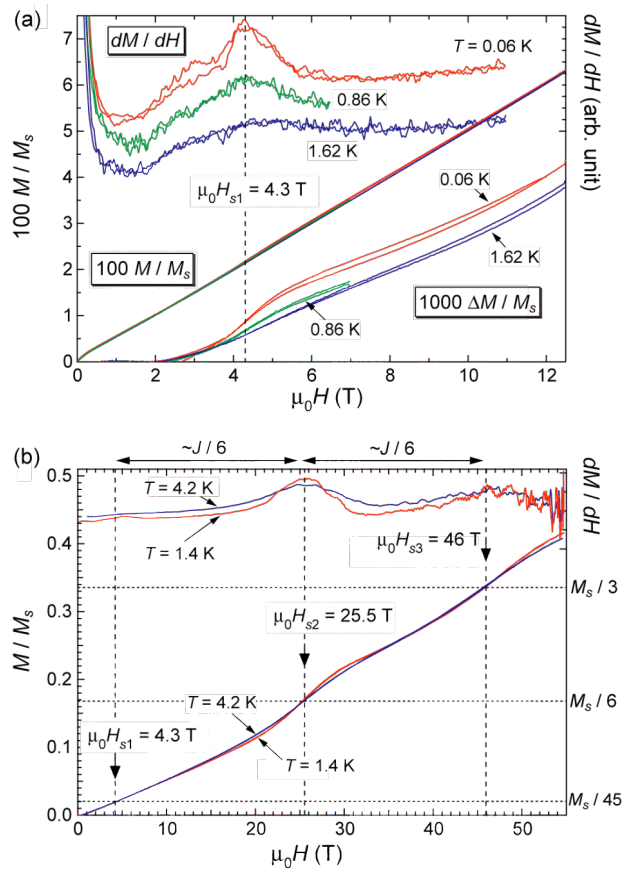

Fig. 3. Magnetization M versus magnetic field H. (a) M divided by the saturation magnetization $M_s$ measured by the Faraday method at $T = 0.06$, 0.86 and 1.62 K at increasing and decreasing fields. The derivative of each curve is shown at the top with appropriate offsets for clarity. The magnetization after the subtraction of the initial linear component and multiplication by a factor of 10 is plotted at the bottom. (b) Magnetization measured using a pulsed magnet up to $\mu_0 H = 55$ T at $T = 1.4$ and 4.2 K. The derivative of each M-H curve is shown at the top. There are three step-like increases at $\mu_0 H_{s1} = 4.3$ T, $\mu_0 H_{s2} = 25.5$ T, and $\mu_0 H_{s3} = 46$ T.

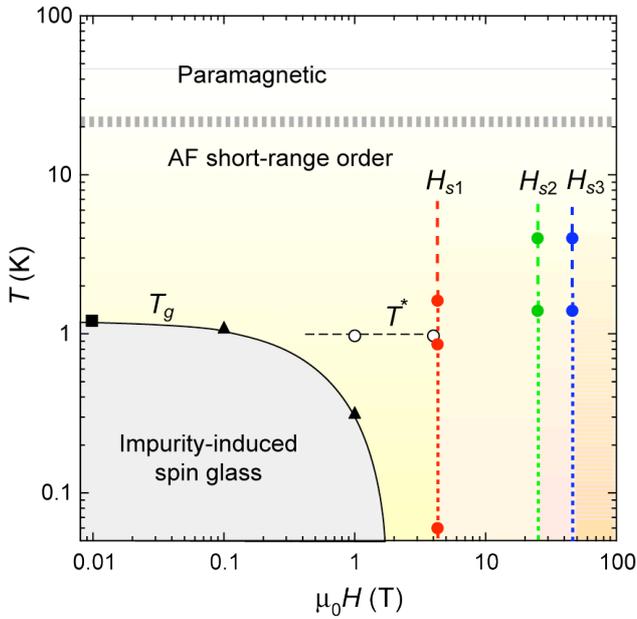

Fig. 4. Phase diagram of volborthite. An antiferromagnetic spin correlation develops below $T_p = 22$ K $\sim J/4$. At low magnetic fields, an impurity-induced spin-glass phase exists with $T_g$ decreasing with increasing $H$ and disappears above 2 T (square and triangle marks indicate results obtained from ref. 23 and the present experiments, respectively). $T^* = 1$ K is the temperature at which a sharp peak in the relaxation rate was observed in NMR experiments. The three nearly vertical lines at $\mu_0 H_{s1} = 4.3$ T, $\mu_0 H_{s2} = 25.5$ T, and $\mu_0 H_{s3} = 46$ T separate the ground

The results presented in this report are summarized as a phase diagram illustrated in Fig. 4. An antiferromagnetic spin correlation develops below $T_p = 22$ K $\sim J/4$, but it cannot grow to a conventional LRO at the lowest temperature. At low magnetic fields, a spin-glass phase exists with $T_g$ decreasing with increasing $H$ and disappears above ~2 T. This spin glass must be induced by impurity spins and is not intrinsic. The true ground state of volborthite may be a gapless spin liquid or a long-range RVB state with a tiny gap, less than $J/1500$. In non-zero magnetic field, there exist at least four spin liquid or other quantum states. A large increase in relaxation rate observed in NMR experiments towards $T^* = 1$ K may not be indicative of a conventional LRO but unusually slow spin dynamics persisting below $T^*$ or a certain hidden order that maintains spin fluctuations down to the lowest temperature. The three magnetization steps separate the ground state into at least four regions. Trying to understand this rich phase diagram would provide us with a chance to come up with a novel idea on the physics of the spin liquid. An interesting physics beyond the standard theoretical predictions must be hidden behind these unexplained phenomena, associated with frustration and quantum fluctuations on the kagome lattice.

We thank J. Yamaura and H. Kawamura for helpful discussion and C. Lhuillier for the naming of "magnetization steps". This work was supported by Grant-in-Aids for Scientific Research on Priority Areas "Novel States of Matter Induced by Frustration" (19052003) and "High Field Spin Science in 100T" (451).